\documentclass[%
 reprint,
 amsmath,amssymb,
 aps,
]{revtex4-2}

\usepackage{graphicx}% Include figure files
\usepackage{dcolumn}% Align table columns on decimal point
\usepackage{bm}% bold math
\begin{document}

%\preprint{APS/123-QED}

\title{Double strangeness molecular-type pentaquarks from coupled channel dynamics}% Force line breaks with \\

\author{J.A. Mars\'e-Valera}
%\affiliation{%
%	Authors' institution and/or address\\
%	This line break forced with \textbackslash\textbackslash
%}%
\author{V.K. Magas}%
% \email{vladimir@fqa.ub.edu}
\author{A. Ramos}
\affiliation{
 Departament de F\'isica Qu\`antica i Astrof\'isica  
 and Institut de Ci\`encies del Cosmos (ICCUB) \\
 Universitat de Barcelona,
 Mart\'i i Franqu\`es 1, 08028 Barcelona, Spain}%


\date{\today}% It is always \today, today,

\begin{abstract}

The existence of pentaquarks with strangeness content zero and one are major discoveries of the latest years in hadron physics. Most of these states can be understood as hadronic molecules and were predicted prior to their discovery within a model based on unitarized meson-baryon amplitudes obtained from vector meson exchange interactions. Contrary to earlier statements, we show this model to also predict the existence of pentaquarks with double strangeness, at about 4500 MeV and 4600 MeV, which are generated in a very specific and unique mechanism, via an attraction induced by a strong coupling between the two heaviest meson-baryon states.

\end{abstract}

\maketitle

%\tableofcontents

%\renewcommand{\figurename}{Fig. }
%\renewcommand{\tablename}{Table }

\section{Introduction} 
\label{sec:int}

Since the turn of the millennium, an increasing amount of data obtained by several collaborations (Belle, BaBar, LHCb, BESII) has produced many exotic hadrons which appear to be inconsistent with the predictions of the
conventional quark model \cite{Godfrey:1985xj,Capstick:1986ter}. The discovery at LHCb \cite{LHCb:2015yax,LHCb:2019kea} of excited nucleon resonances $( P^N_{\Psi s}(4312)$, $P^N_{\Psi s}(4440)$ and $P^N_{\Psi s}(4457))$, seen on the invariant mass distribution of $J/\Psi\, p$ pairs from the decay $\Lambda_b \to J/\Psi\ p\ K^-$ , has been the first clear evidence of the existence of pentaquark baryons, as a $c\bar{c}$ pair is necessary to explain their high mass. In a later experiment \cite{LHCb:2020jpq}, the $J/\Psi\, \Lambda$ spectrum, obtained from the decay $\Xi^-_b \to J/\Psi\ \Lambda\ K^-$, also provided evidence for a pentaquark, the  $P^\Lambda_{\Psi s}(4459)$, this time with one unit of strangeness content. More recently,  an analysis of the $B^- \to J/\Psi \ \Lambda\ {\bar p}$ decay points at the existence of another narrow hidden-charm pentaquark with strangeness,  the  $P^\Lambda_{\Psi s}(4338)$ \cite{Penta4338}. 

Among the various theoretical interpretations on the nature of these pentaquarks, the possibility that they could be structured as meson-baryon molecules was already predicted in \cite{Hofmann:2005sw,Wu:2010jy,Wu:2010vk}, prior to their discovery, and has gained interest ever since due to their proximity to various charmed meson-baryon thresholds (see \cite{Karliner:2022erb,Dong:2021bvy} and references therein). Many of these studies assume a t-channel vector meson exchange interaction between mesons and baryons, 
%obtained from the hidden gauge formalism and 
which is unitarized in coupled-channels. However, a hidden charm pentaquark with double strangeness was not found in this approach \cite{Wu:2010vk,Wu:2010jy}. 
From a hadronic molecular perspective, it is interesting to note that the usually adopted alternative of molecular binding through a long-range interaction mediated by the exchange of pions between the constituent hadrons is not possible in the heavy part of strangeness $S=-2$ sector. In fact, a (no-pion) light meson exchange model  \cite{Wang:2020bjt} finds doubly-strange hidden charm pentaquarks, but requiring a unrealistically large regularization cut-off parameter. Let us finally mention that recent studies relying on the quark substructure of the state also predict double strangeness pentaquark candidates \cite{Ferretti:2020ewe,Azizi:2021pbh,Ortega:2022uyu}.

%The strangeness $S=-2$ is an special sector because it does not allow a long-ranged interaction mediated by the exchange of pions between the constituent hadrons
%, which might be the reason why its possible existence has hardly been discussed in the literature. 
%A recent study \cite{Wang:2020bjt}, based on a light meson exchange potential, finds candidates for doubly-strange hidden charm pentaquarks, in the form of $\Xi^\prime_c D^*_s$  or $\Xi^*_c D^*_s$ molecules, at energies slightly above 4.5 GeV, albeit employing regularization cutoff values larger than 2 GeV, which deviates considerably from the expected value of 1 GeV representing a typical hadronic scale. 

%The unitarized coupled-channel hidden gauge formalism has been a very successful approach in explaining other exotic hadrons first in the strange, and later in the charm and hidden charm sectors (see \cite{Guo:2017jvc} and references therein). The most known example is the $\Lambda(1405)$ resonance, whose mass was systematically predicted to be too high by quark models, but it found a better explanation as a meson-baryon bound state. Furthermore, the chiral models with unitarization in the coupled channels have predicted its double-pole nature~\cite{Oller:2000fj,Jido:2003cb}, which can be seen from comparing different experimental line shapes~\cite{Magas:2005vu}, and which is now commonly accepted in the field and appears in the PDG ~\cite{PDG}.

The unitarized coupled-channel hidden gauge formalism has been a very successful approach in explaining other exotic hadrons first in the strange, and later in the charm and hidden charm sectors (see \cite{Guo:2017jvc} and references therein). The most known example is the $\Lambda(1405)$ resonance, whose mass was systematically predicted to be too high by quark models, but it found a better explanation as a resonant structure from a coupled-channel ${\bar K}N$ interaction. This picture, which would consolidate the $\Lambda(1405)$ as the first pentaquark ever, finds support in its
predicted double-pole nature~\cite{Oller:2000fj,Jido:2003cb}, confirmed later from the different line shapes observed~\cite{Magas:2005vu}, and now commonly accepted in the field~\cite{PDG}.

%In this work, we revisit the procedure of the unitarized coupled-channel hidden gauge formalism by relaxing some of the approximations carried out in \cite{Wu:2010vk,Wu:2010jy} and still employing realistic regularization parameters. And we demonstrate that the model does generate strangeness $S=-2$ pentaquarks via a non-diagonal attraction between the two heaviest meson-baryon channels. For our knowledge, this is the first situation in hadron physics there such a mechnism not only influences the position and width of the molecular state, but really determines the existence or non-existence of the resonance. 

In this work, we revisit the unitarized coupled-channel approach \cite{Wu:2010vk,Wu:2010jy} and demonstrate that, within realistic model parameters,  it does predict the existence of strangeness $S=-2$ pentaquarks in a very unique way, namely via an attraction generated by a strong coupling between the two heaviest meson-baryon channels. This finding not only complements the family of $S=0$, and $S=-1$ pentaquarks obtained within the same formalism but, if confirmed experimentally, it would strongly support the validity of the unitarized t-channel vector-meson exchange interaction models in explaining hadron molecules, as the alternative pion-exchange mechanism is forbidden in this sector.
%For our knowledge, this is the first situation in hadron physics there such a mechanism not only influences the position and width of the molecular state, but really determines the existence or non-existence of the resonance. 

\section{Formalism}
\label{sec:formalism}
%The model of the meson-baryon interaction used in this work is based on the tree-level diagrams of Fig. \ref{Fig:treediagr}. In this work we will only consider the s-wave amplitude, where the most important contribution is the t-channel term (Fig. \ref{Fig:treediagr}(a)). The s- and u- channel terms (Fig. \ref{Fig:treediagr}(b) and (c)) contribute mostly to the p-wave amplitude and they may have effects at higher energies. In Ref. \cite{Oller:2000fj} it was shown that the contribution from these terms in the sector $S=-1$ can reach up to $20\%$ of the dominant t-channel contribution, while in the sector $S=-2$, since the intermediate baryons are more massive, we can that them to be even less important and therefore we will neglect its contribution.

\begin{figure}[h!]
     \centering
     \includegraphics[scale=0.40]{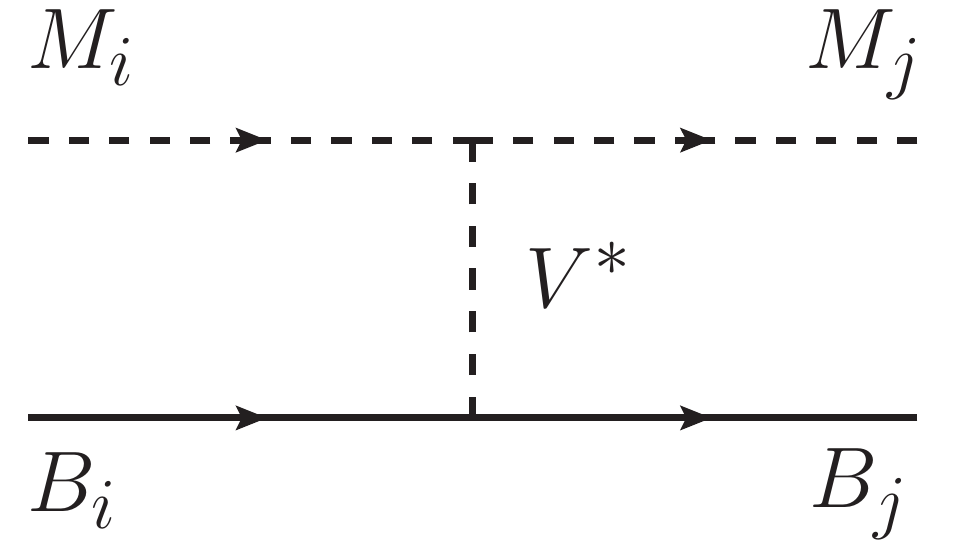}
     \caption{Leading order tree level diagram contributing to the meson-baryon interaction. Baryons and mesons are depicted by solid and dashed lines, respectively.}
     \label{Fig:treediagr}
 \end{figure}

The model employed in this work relies on a meson-baryon interaction in S-wave, which is built up from the t-channel diagram of Fig.~\ref{Fig:treediagr}, where the mesons are depicted by dashed lines and the baryons by solid lines. The external mesons in the diagram represent the ground states of either pseudoscalar ($J
^P(M)=0^-$)- or vector ($J
^P(M)=1^-$)-type, while the baryons are the lowest-energy ones with $J^P(B)=1/2^+$, all of them composed out of $u,d,s,c$ quarks and antiquarks. The indices $i,j$ label the different meson-baryon channels with the same spin-parity ($J^P$), isospin and flavor quantum numbers that can be connected by the exchange of a vector meson $V^*$.

The $VPP$ ($VVV$) and $VBB$ vertices in the diagram are described by effective Lagrangians which are obtained using the local hidden gauge formalism \cite{Bando:1984ej,Bando:1987br} and assuming SU(4) symmetry: 
\begin{eqnarray}\label{eq:vertices}
\mathcal{L}_{VPP}&=&-{\rm i} g\langle\left[\phi, \partial_\mu\phi\right] V^\mu\rangle \ ,\nonumber \\
%\end{equation}
%\begin{equation}\label{eq:vertexBBV}
\mathcal{L}_{VBB}&=&\frac{g}{2}\sum_{i,j,k,l=1}^4\bar{B}_{ijk}\gamma^\mu\left(V_{\mu,l}^{k}B^{ijl}+2V_{\mu,l}^{j}B^{ilk}\right)\!\!\ , \\
\mathcal{L}_{VVV}&=&{\rm i} g\langle {\left[V^\mu,\partial_\nu V_\mu\right] V^\nu}\rangle \ , \nonumber
\end{eqnarray}
where $\phi$ and $V_\mu$ represent the 16-plet of pseudoscalar and vector fields, respectively, $\langle~\rangle$ denotes the trace in $SU(4)$ flavour space, and $g=m_V/2f$ is the universal coupling constant, related to the pion decay constant $(f=93~{\rm MeV})$ via a characteristic mass of the light uncharmed vector meson from the nonet $(m_V)$.

%To describe the interaction we employ the effective Lagrangians that describe the vertex couplings of the vector meson to pseudoscalar mesons $(VPP)$ and baryons $(VBB)$ which are obtained using the hidden gauge formalism and assuming $SU(4)$ symmetry \cite{Hofmann:2005sw}:
%\begin{equation}\label{eq:vertexVPP}
%\mathcal{L}_{VPP}=ig\langle\left[\partial_\mu\phi, \phi\right] V^\mu\rangle,
%\end{equation}
%\begin{equation}\label{eq:vertexBBV}
%\mathcal{L}_{VBB}=\frac{g}{2}\sum_{i,j,k,l=1}^4\bar{B}_{ijk}\gamma^\mu\left(V_{\mu,l}^{k}B^{ijl}+2V_{\mu,l}^{j}B^{ilk}\right),
%\end{equation}
%where $\phi$ and $V_\mu$ represents the pseudo-scalar field of 16-plet and the vector field of 16-plet, %$\langle~\rangle$ denotes the $SU(4)$ trace in flavour space. The $g=m_V/2f$ is a coupling constant, %related to the pion decay constant (experimental value $f=93~{\rm MeV})$) and a characteristic mass of %a light vector meson from nonet $(m_V)$.

Using the $VPP$ and $VBB$ Lagrangians of Eq.~(\ref{eq:vertices}) one can obtain the interaction of pseudoscalar mesons with baryons, referred to as $PB$ interaction from now on.  In the limit
where the mass of the exchanged meson is much bigger than its four-momentum,  the t-channel 
diagram of Fig.~\ref{Fig:treediagr} reduces to a contact term and, up to $\mathcal{O}(p^2/M^2)$, its S-wave expression reads:
\begin{equation}\label{eq:Vij}
	V_{ij}(\sqrt{s})=-C_{ij}\frac{1}{4f^2}\left(2\sqrt{s}-M_i-M_j\right) N_i N_j,
\end{equation}
where $N_i$ and $N_j$ are normalization factors, $N_i=\sqrt{(E_i+M_i)/2M_i}$, and $M_i$ ($E_i$), $M_j$ ($E_j$) are the mass (energy) of the baryon in the incoming and outgoing channels, respectively.  
%Despite of the $C_{ij}^v$ is $SU(4)$ symmetrical, since we use the physical masses of the mesons and baryons our interaction potential is not $SU(4)$ symmetric.
The coefficients $C_{ij}$ contain the $SU(4)$ Clebsh-Gordan factors encoded in the Lagrangians of Eq.~(\ref{eq:vertices}), as well as the ratio between $m_V^2$ and the mass squared of the vector meson exchanged which, in the $SU(4)$ limit, is also $m_V^2$. However, the transitions studied here involve the exchange of vector mesons with quite different masses and this is taken into account as follows. The mass of all the light vector mesons, composed out of $u,d,s$ quarks and their corresponding antiquarks, is assumed to be the same and equal to $m_V$. The mass of mesons with one charm or anticharm quark is assumed to be roughly twice that of the uncharmed ones, hence the coefficients tied to their exchange will carry a factor $\kappa_c=(m_V/m_V^c)^2\simeq 1/4$. Finally, the mass of the $J/\Psi$ meson is taken to be three times that of the light ones, so that the terms induced by its exchange carry a factor  $\kappa_{cc}\simeq 1/9$ in this case.

In the present work we focus on the isospin $I=1/2$ and strangeness $S=-2$ sector of the pentaquark, which can be realized by nine possible combinations of a pseudoscalar meson and a baryon, namely $\pi\Xi(1456)$, $\bar{K}\Lambda(1611)$, $\bar{K}\Sigma(1689)$, $\eta\Xi(1866)$, $\eta'\Xi(2276)$, $\eta_c\Xi(4298)$, $\bar{D}_s\Xi_c(4437)$, $\bar{D}_s\Xi_c^\prime(4545)$, and $\bar{D}\Omega(4565)$, where the values in parenthesis indicate their thresholds in MeV. As the mass of the four channels with charm quarks is more than $2~{\rm GeV}$ larger than that of the other channels, we can then expect these two energy regions to behave independently and the corresponding channels to have a small influence on each other. Hence, we consider only the four heavy channels and present the corresponding values of the $C_{ij}$ coefficients in Table \ref{Tab:CijPB}. 
\begin{table}[h!]
  \begin{center}
    \begin{tabular}{c|ccccc}
        \hline\\ [-0.30cm]
         & $\eta_c\Xi$ & $\bar{D}_s\Xi_c$ & $\bar{D}_s\Xi_c^\prime$ & $\bar{D}\Omega_c$\\
        \hline \\ [-0.2cm]
        $\eta_c\Xi$ & $0$ & $\sqrt{\frac{3}{2}}\kappa_c$ & $\frac{1}{\sqrt{2}}\kappa_c$ & $-\kappa_c$\\ 
        $\bar{D}_s\Xi_c$ &  & $-1+\kappa_{cc}$ & $0$ & $0$\\
        $\bar{D}_s\Xi_c^\prime$ & & & $-1+\kappa_{cc}$ & $\sqrt{2}$\\
        $\bar{D}\Omega_c$ & & & & $\kappa_{cc}$ \\
    \end{tabular}
  \end{center}
\caption{$C_{ij}$ coefficients of the $PB$ the interaction in the $(I,S)=(1/2,-2)$ sector.}    
\label{Tab:CijPB}
\end{table}

Similarly, using the $VVV$ and $VBB$ Lagrangians of Eq.~(\ref{eq:vertices}) one can obtain the interaction of vector mesons with baryons, referred to as the $VB$ interaction from now on. Neglecting the momentum of the external vector mesons versus their mass, their polarization vectors become essentially spatial and the resulting $VB$ interaction kernel is then identical to that for the interaction of pseudoscalar mesons with baryons, Eq.~(\ref{eq:Vij}), but multiplied by the scalar product $\Vec{\epsilon}_i\cdot\Vec{\epsilon}_j$. 

%One can see that in this case we can arrive to a similar expression from Eq. (\ref{eq:Vij}) but multiplied by  the product of the  polarization vectors $\Vec{\epsilon}_i\cdot\Vec{\epsilon}_j$.
%\begin{equation}\label{eq:VijVB}
% V_{ij}(\sqrt{s})=-C_{ij}\Vec{\epsilon}_i\cdot\Vec{\epsilon}_j\frac{1}{4f^2}\left(2\sqrt{s}-M_i-M_j\right) N_i N_j,
%\end{equation}

In the $VB$ interaction case, the allowed channels are $\rho\Xi(2089)$, $\bar{K}^*\Lambda(2010)$, $\bar{K}^*\Sigma(2087)$, $\omega\Xi(2101)$, $\phi\Xi(2338)$, $J/\Psi\,\Xi(4415)$, $\bar{D}^*_s\Xi_c(4581)$, $\bar{D}^*_s\Xi_c^\prime(4689)$, and $\bar{D}^*\Omega(4706)$, where again we can separate the channels in two energy regions. Focusing on the heavy channels, the matrix of $C_{ij}$ coefficients is identical to that in Table~\ref{Tab:CijPB} with the following replacements in the labeling:  $\eta_c \to J/\Psi$, $\bar{D}\rightarrow\bar{D}^\ast$, $\bar{D}_s\rightarrow\bar{D}_s^\ast$.

The sought resonances are dynamically generated as poles of a properly unitarized scattering amplitude $T_{ij}$. We employ the Bethe-Salpeter equation in coupled channels, which implements a re-summation of meson-baryon loops, denoted by $G_l$, to infinite order:
\begin{equation}\label{eq:Tij1}
T_{ij}=V_{ij}+V_{il}G_lT_{lj}
\end{equation}
Note that, in the case of the $VB$ amplitude, the sum over the polarization of the internal vector meson gives 
$\sum_{pol}\epsilon_i\epsilon_j=\delta_{ij}+\frac{q_iq_j}{M_V^2} $
but, consistently with our model, the $q^2/M_V^2$ term is neglected and this permits factorizing a factor $\Vec{\epsilon}_i\Vec{\epsilon}_j$ out from all terms in the Bethe-Salpeter equation.
By adopting the commonly employed approximation of factorizing the $V$ and $T$ on-shell amplitudes out of the intermediate integrals, the above integral equation reduces to a simple algebraic one,
%\begin{equation}\label{eq:Tij2}
   $ T=(1-VG)^{-1}V \ ,$
%\end{equation}
where 
$G$ is a diagonal matrix containing the meson-baryon loop functions given by
\begin{equation}\label{eq:Gl}
G_l(P)=i\int\frac{d^4q}{(2\pi)^4}\frac{2 M_l}{(P-q)^2-M_l^2+i\epsilon}\frac{1}{q^2-m_l^2+i\epsilon} \ ,
\end{equation}
$M_l$ and $m_l$ denoting the mass of the baryon and meson in the loop, respectively, $P=p+k=(\sqrt{s},0)$ being the total four-momentum in the c.m. frame, and $q$  the internal meson four-momentum. The loop function diverges logarithmically and needs to be regularized. We employ the {\it cutoff  method} that implies restricting the momentum integrals up to a value $\Lambda$, which should be related to the dynamical scale that has been integrated out in our model, like the mass of the lighter vector mesons being exchanged in the t-channel diagram. Taking these considerations into account, we choose $\Lambda=800$ MeV.
%Alternatively, one can employ {\it dimensional regularization} techniques, which implies the appearance of a set of loop subtraction constants that also need to be mapped out to some scale. 

The resonances are generated as poles of the scattering amplitude $T_{ij}$, analytically continued to the so-called \textit{second Riemann sheet} of the complex energy plane. The behavior of the amplitude in the vicinity of the pole allows one to obtain the couplings $g_i$ of the resonance to the various meson-baryon channels, as well as the compositeness $\chi_i$, which measures the amount of a given meson-baryon component in the resonance. See, for example, Ref.~\cite{Montana:2017kjw} for more details on the formalism and explicit definitions.

\section{Results and Discussion}
\label{sec:res}

Following our model, in the $PB$ sector we obtain  a dynamically generated hidden charm double-strange baryon resonance at $4493$ MeV with a width of  $74$ MeV. As can be seen from the detailed information given in Table~\ref{tab:PBHigh}, this resonance couples strongly to the $\bar{D}\Omega_c$ and $\bar{D}_s\Xi_c^\prime$ channels and is dominantly composed by  $\bar{D}\Omega_c$ states. Being generated by the S-wave interaction of a $0^-$ pseudoscalar meson with a $1/2^+$ baryon in the $(I,S)=(1/2,-2)$ sector, the $P^\Xi_{\Psi ss}(4493)$ state is predicted to have spin-parity $J^\pi=1/2^-$.
 
 \begin{table}[hbt!]
	\begin{center}
		\begin{tabular}{ccccc}
			\hline\\
			\multicolumn{5}{l}{$0^- \oplus 1/2^+$ $PB$ interaction in the $(I,S)=(1/2,-2)$ sector}\\[2mm]
			\hline\\ 
			$P^\Xi_{\Psi ss}(4493)$ & & \!\!\!\!\!\!\! {$M_R=4493.35$\ MeV} & \multicolumn{2}{c}{~~
			$\Gamma_R=73.67$\ MeV}\\
			\hline
			& & \multicolumn{2}{c}{} & \\ [-3mm]
			& ${\rm thr.}$ (MeV) & $g_i$ & $|g_i|$ & $ \chi_i$ \\
			$\eta_c \Xi$            &  4298 & $-1.60+{\rm i\,} 0.34$ &\ 1.63 & 0.220\\
			$\bar{D}_s \Xi_c$     &  4437 & $-0.17+{\rm i\,}0.27$ &\ 0.32 & 0.019\\
			$\bar{D}_s \Xi_c^\prime$    &  4545 & $ -2.41+{\rm i\,}0.58$ &\ 2.48 & 0.398\\
			$\bar{D}\Omega_c$    & 4564 & $ 3.59-{\rm i\,}0.77$ &\ 3.67 & 0.711\\
			\hline
		\end{tabular}
	\end{center}
	\caption{Energy, width, couplings to meson-baryon states, and compositeness of the $PB$ molecular $P^\Xi_{\Psi ss}(4493)$ state.}
	\label{tab:PBHigh}
\end{table}
 
We observe that the $P^\Xi_{\Psi ss}(4493)$ appears below the thresholds of the $\bar{D}\Omega_c$ and $\bar{D}_s\Xi_c^\prime$ channels, which participate strongly in its generation, and can only decay into the two lightest ones. It is clear, however, that the most probable decay process will be $P^\Xi_{\Psi ss}(4493)\to \eta_c\Xi$, as the coupling to this lowest energy channel is almost an order of magnitude larger than that to $\bar{D}_s\Xi_c$, together with the fact that the available phase space is also much larger.  Therefore, this state could be seen from the invariant mass spectrum of  $\eta_c\ \Xi$ pairs in the decays $\Xi_b \to \Xi\ \eta_c\ \phi$ and $\Omega_b \to \Xi\ \eta_c \ {\bar K}$.

It is interesting to compare our results with those of other works using similar models.  First of all, we note that a dynamically generated state, strongly coupled to the $\bar{D}\Omega_c$ and $\bar{D}_s\Xi_c^\prime$ channels, is also found in Ref. \cite{Hofmann:2005sw}, but at a much smaller energy,  
$M_R\simeq 3800$~MeV. 
%This state is below the thresholds of all the heavy channels, and thus it can only decay via the light channels of this sector, to which it couples very weakly. Therefore its very narrow width $\Gamma_R\simeq 6$ MeV is not surprising. 
Such a substantial difference has to do with the rather different method and parameters of the regularization scheme employed.  While we use the {\it cutoff  method} with $\Lambda=800$~MeV, the authors of Ref. \cite{Hofmann:2005sw} regularize the loop function imposing the requirement $G_l(\mu)=0$, at $\mu$ given by $\mu=\sqrt{m_{\rm th}^2+M_{\rm th}^2}$, being $m_{\rm th}$ and $M_{\rm th}$ the masses of the meson and baryon, respectively, of the lightest possible channel, namely $\pi\, \Xi$  in the case of Ref. \cite{Hofmann:2005sw}. Enforcing this regularization requirement amounts to using in our model of a cutoff value of $\Lambda\simeq 2800~{\rm MeV}$, which is unreasonably large in our opinion.

We next compare our results with those of  Ref. \cite{Wu:2010vk}, where the loop function is calculated in a way similar to ours, but the resonance strongly coupled to the $\bar{D}\Omega_c$ and $\bar{D}_s\Xi_c^\prime$ states is not discussed, possibly because the mild attraction induced by $J/\Psi$ exchange is neglected taking $C_{44}=0$ (which in our case is equivalent to setting $\kappa_{cc}=0$) and the remaining diagonal channels have null or repulsive interactions. Ignoring  the small value of $\kappa_{cc}$ is usually a sufficiently good approximation. In fact, the mild attractive character of $C_{44}=\kappa_{cc}=1/9$ would not be sufficient to produce a $\bar{D}\Omega_c$ bound state in an uncoupled channels calculation, as we have numerically checked. However, in this particular sector, one finds a sizable non-diagonal coefficient, $C_{43}=-\sqrt{2}$, which, via coupled channels, adds enough attraction to make the dynamical generation of the new molecular state possible. We have tested that the minimum strength of $C_{43} $ which can still generate this resonance is $-0.75\sqrt{2}$. We are thus observing a very interesting phenomenon, uniquely related to the coupled channel dynamics, which enhances the mild attraction of the $\bar{D}\Omega_c$ interaction to produce a bound state. This is analogous to what happens to the deuteron, which appears as a bound state of the coupled-channel $^3S_1-^3D_1$ nucleon-nucleon interaction, but would disappear in an uncoupled calculation \cite{MACHLEIDT19871}.

\begin{figure}[h!]
	\centering
	\includegraphics[scale=0.5]{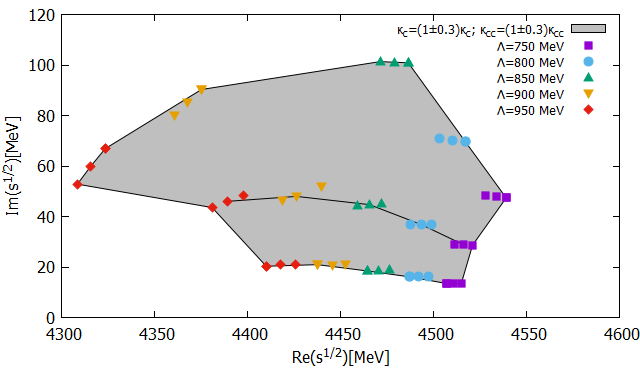}
	\caption{Region in the complex energy plane where the mole\-cular $P^\Xi_{\Psi ss}$ pentaquark can be found. The coloured symbols indicate the pole position of the resonance for different cutoff values in the range $(750-950)$ MeV and
nine different  $SU(4)$-breaking parameters,  corresponding  to  the  possible  combinations  of  the values $(0.7, 1.0, 1.3)\kappa_c$ and $(0.7, 1.0, 1.3)\kappa_{cc}$.  The upper dots are obtained with $1.3\kappa_c$ and the lower ones  with  $0.7\kappa_c$. The dependence on the variation of $\kappa_{cc}$ is mild and is reflected in the separation within each 3-symbol group. The  black  lines  are  merely  to  guide  the  eye  and  join  the  results  obtained  with different cutoff values for the combinations with $1.3\kappa_c, 0.7\kappa_{cc}$  (upper line), $\kappa_c,\kappa_{cc}$ (middle line) and $0.7\kappa_c, 1.3\kappa_{cc}$  (lower line).
}
	\label{fig:SU4}
\end{figure}

While our model relies on SU(4) symmetry, it is already violated to some extent by the use of the physical masses for the mesons and baryons in the interaction kernel and in the loop function. However, we would like to analyze how the generated resonance is affected when we introduce up to an additional 30\% of $SU(4)$ symmetry violation in the coupling vertices of the transitions mediated by the exchange of heavy mesons. This is implemented in practice by multiplying $\kappa_c$ and $\kappa_{cc}$ by a factor varying in the range $(0.7 - 1.3)$. We also investigate the sensitivity of the pole position to the cutoff parameter $\Lambda$, which we vary from $750$ MeV until $950$ MeV.
The results of all these calculations form the gray area displayed in Fig. \ref{fig:SU4}.
% which represents the region in the complex energy plane where the resonance can be found, assuming up to 30\% of $SU(4)$ breaking in the heavy-meson mediated transitions and cutoff values in the range $(750-950)$~MeV. 
It is seen that modifications of the cutoff produce sizable variations in the mass of the resonance but a moderate effect on its width. In contrast, a modification of $\kappa_c$ produces important variations in the width of the resonance, while the mass is less affected, except for the larger cutoff and $\kappa_c$ values. This is easily understood from inspecting the coefficients of Table~\ref{Tab:CijPB}, which show that the size of the parameter $\kappa_c$ is directly affecting the couplings of the heavier channels to the lighter one, $\eta_c\Xi$, to which the resonance decays. The overall conclusion is that, although the mass and the width depend on the model parameters, the appearance of a pole is a robust outcome in all these calculations, giving support to our claim of the probable existence of a $(I,S)=(1/2,-2)$ molecular-type pentaquark $P^\Xi_{\Psi ss}$ with spin-parity $J^P=1/2^-$.

 \begin{table}[hbt!]
	\begin{center}
		\begin{tabular}{ccccc}
			\hline\\
			\multicolumn{5}{l}{$1^- \oplus 1/2^+$ $VB$ interaction in the $(I,S)=(1/2,-2)$ sector}\\[2mm]
			\hline\\ 
			$P^\Xi_{\Psi ss}(4633)$ & & \!\!\!\!\!\!\! {$M_R=4633.38$\ MeV} & \multicolumn{2}{c}{~~
			$\Gamma_R=79.58$\ MeV} \\
			\hline
			& & \multicolumn{2}{c}{} & \\ [-3mm]
			& ${\rm thr.}$ (MeV) & $g$ & $|g_i|$ & $ \chi_i$ \\
			$J/\psi\,\Xi$            &  4415 & $-1.62+{\rm i\,}0.38$  & 1.66 & 0.252\\
			$\bar{D}_s^*\Xi_c$     &  4581 & $-0.143+{\rm i\,}0.32$ & 0.34 & 0.022\\
			$\bar{D}_s^*\Xi_c^\prime$    &  4689 & $-2.49+{\rm i\,}0.67$  & 2.58 & 0.406\\
			$\bar{D}^*\Omega_c$    &  4706 & $3.67+{\rm i\,}0.89$   & 3.78 & 0.740\\
			\hline
		\end{tabular}
	\end{center}
	\caption{Energy, width, couplings to meson-baryon states, and compositeness of the $VB$ molecular $\Xi(4633)$ state.}
	\label{tab:VBHigh}
\end{table}

Proceeding in a similar way, we obtain in the $VB$ sector a resonance at $4633$ MeV with a width of around $80$ MeV, in agreement with \cite{Azizi:2021pbh}, strongly coupled to the $\bar{D}^*\Omega_c$ and $\bar{D}_s^*\Xi_c^\prime$ channels, dominantly composed by $\bar{D}^*\Omega_c$ states, and decaying mostly to  $J/\Psi\, \Xi$ pairs. The detailed values of the pole position, couplings and compositeness can be found in Table \ref{tab:VBHigh}. Note that this $P^\Xi_{\Psi ss}(4633)$ state is degenerate in spin since, having been obtained from the interaction in S-wave of $J^P=1^-$ vector mesons with  $J^P=1/2^+$ baryons, it can have either $J^P=1/2^-$ or $J^P=3/2^-$. Similarly to the $PB$ resonance discussed before, the  $P^\Xi_{\Psi ss}(4633)$ pentaquark could be seen as a peak in the invariant mass spectrum of $J/\Psi\, \Xi$ pairs produced in the decays $\Xi_b \to \Xi\ J/\Psi\  \phi$ and $\Omega_b \to \Xi\ J/\Psi \ {\bar K}$.

\section{Conclusions}
\label{sec:conclusions}
Stimulated by the recent discoveries by the LHCb collaboration of hidden charm pentaquark states with strangeness $S=0$ and $S=-1$, some of which could be associated to the predicted meson-baryon molecular states found in unitary models based on vector-meson exchange interactions, we have revisited these models to study the possible existence of pentaquarks with strangeness $S=-2$.

Employing realistic regularization parameters, we predict double strangeness pentaquarks of molecular nature around 4500 MeV and 4600 MeV. This unexpected result of the model was overlooked before, due to the dominantly repulsive interaction between the meson and the baryon in various channels of the basis employed. However, we have found that bound states are indeed generated in a very specific and unique way, via a strong non-diagonal attraction between the two heaviest meson-baryon channels. 
%This effect was overlooked before
%because the very mild attraction in the diagonal interaction of the heaviest meson-baryon state was neglected, hence effectively preventing the %coupled-channel mechanism to take effect. 

Our work should stimulate experiments looking for these type of pentaquarks, the discovery of which would enrich the family of their already observed $S=0$ and $S=-1$ pentaquark partners. The absence in this sector of a long range interaction mediated by pion-exchange also makes the search for these states specially interesting. If they do exist, their interpretation as molecules would require a change of paradigm, since they could only be bound through heavier-meson exchange models as the one employed in this work.
 
% < Reply 
%It is interesting to note that in our model the pion exchange between interacting meson and baryon is not possible, in contrast, for example, to the light meson exchange potential model  \cite{Wang:2020bjt}. If such states really do exist, they will require a major paradigm shift, because in all molecular pentaquarks, observed so far, the pion exchange is possible, and, in complete analogy with nuclear physics, it is thought to be responsible for the long range part of the interacting potential. 
% > Reply 

%Our work should stimulate experiments looking for these type of pentaquarks, the discovery of which would enrich the family of their already observed $S=0$ and $S=-1$ pentaquark partners. 
%The first one, $\Xi(4493)$, is a $J^P=1/2^-$ resonance generated from the interaction of pseudoscalar mesons with baryons, which couples most strongly to the $\bar{D}\Omega_c$ and $\bar{D}_s\Xi_c^\prime$ channels and can be most likely observed via its decay to $\eta_c \Xi$ states, for instance in decay processes of bottomed baryons, such as $\Xi_b \to \Xi\ \eta_c\ \phi$ and $\Omega_b \to \Xi\ \eta_c\ {\bar K}$.  The other one, $\Xi(4633)$, is a spin-degenerate resonance which can have  $J^P=1/2^-$ or $J^P= 3/2^-$ and is obtained from the interaction of vector mesons with baryons. It couples dominantly to the $\bar{D}^*\Omega_c$ and $\bar{D}_s^*\Xi_c^\prime$ channels and  it should be looked for in invariant mass spectra of $J/\Psi\, \Xi$ pairs which could be produced in the decays $\Xi_b \to \Xi\ J/\Psi\ \phi$ and  $\Omega_b \to \Xi\ J/\Psi\ {\bar K}$.

\section{Aknowledgements} 
%{\bf Aknowledgements.\ \  }
V.K.M. and A.R. acknowledge support from the Ministerio de Ciencia e Innovaci\'on of Spain through the “Unit of Excellence María de Maeztu 2020-2023” award to the Institute of Cosmos Sciences (CEX2019-000918-M) and under~the project PID2020-118758GB-I00. Support from the EU STRONG-2020 project under the program H2020-INFRAIA-2018-1, grant agreement no.\ 824093, is also acknowledged.

%\nocite{*}

\bibliography{apssamp}% Produces the bibliography via BibTeX.

\end{document}